\documentclass[12pt,preprint,showkeys]{revtex4-2}
\usepackage{graphicx}
\usepackage{amsmath}
\usepackage{amsfonts,amssymb}
\usepackage[matrix,arrow,curve,frame,poly,arc]{xy}
\usepackage[english]{babel}

\def\be{\begin{equation}}
\def\ee{\end{equation}}
\def\ba{\begin{eqnarray}}
\def\ea{\end{eqnarray}}

\begin{document}
\bibliographystyle{plainnat}

\title{Some exact solutions of Friedmann cosmological equation}

\author{Maria Shubina}




\begin{abstract}

In this paper we present a number of examples of exact solutions for the Friedmann cosmological equation for metric $ F(R) $ gravity model. Emphasis was placed on the possibility of obtaining exact time dependences of the main cosmological physical quantities: scale factor, scalar curvature, Hubble rate and function $ F(R) $. For this purpose an ansatz was used to reduce the Friedmann equation to an ordinary differential equation for function $ F = F(H^{2})$. This made it possible to obtain a number of exact solutions, both already known and new. 
\end{abstract}


\maketitle
\newpage

\section{Introduction}

From the first quarter of the last century to this day General Theory of Relativity (GR) remains the best reliable and consistent fundamental theory describing gravitational interaction. However working perfectly on the scale of the Solar System and reproducing Newtonian gravity, General Theory of Relativity cannot provide a unified and theoretically consistent description of the various evolution eras of the Universe; in particular, it is unable to explain the epoch of acceleration in the early stages of evolution known as inflation and the effects of late acceleration associated with the presence of dark energy. Therefore, the problem arose to modify General Relativity in order to create a gravity model that could describe the various evolution stages of our Universe. The criterion for the viability of a modified theory of gravity, like most physical theories, is the correspondence of theoretical predictions to observations. So far observations of events occurring in the Universe and available to us mainly consist of large-scale events and observations associated with compact gravitational objects. Therefore one can expect that the accumulation of experimental data not explained by GR will contribute to the development of modified gravity theories that go beyond the framework of GR and the Standard Model of particle physics. General Relativity is based on the Einstein-Hilbert gravitational action with Lagrange density $ \sqrt{-g} R $, where $ R $ is the Ricci curvature ($ g = \det g_{\mu \nu }$), and among the modified gravity theories the most popular is the one in which the Lagrangian is considered as function $F(R)$, see \cite{SF_2010}-\cite{NOO} and references therein. The various forms of $F(R)$ have appeared in the literature over the past few decades. Among these functions there are quite viable ones that correctly describe the cosmological dynamic, smooth transition between different cosmological eras, correct weak-field limit and dynamics of cosmological perturbations \cite{F2011}. For a flat space-time solution to exist the authors consider the functions $F(R)$ as $ F(R) \sim R + f(R) $ where $ f(R) $ may have a rather arbitrary form, in particular $ f(R) \sim (R + C)^k$, $ f(R) \sim e^{CR} $ and others \cite{NO_2003} - \cite{OO_2021}. In the context of viable models capable of describing both the accelerating expansion of the Universe in our time and in early times, one can formulate a number of conditions on the function $ f(R) $, for example $\lim_{R \to \infty} f(R) = - \Lambda $ \cite{NOO}. Also many authors consider such functions $F(R)$ that do not give Minkowski space-time in the limit, for example $ F(R) \sim (R + C)^k$ where $ k $ can be any number including very close to one \cite{F2011}, \cite{NO_2003}, \cite{NO_2008}, \cite{CCT_2003} - \cite{S_2023}. Since $F(R)$ gravity is "the most sound representative theory of modified gravity" \cite{NOO}, it is interesting to find different possible solutions to this function and to study whether these solutions admit of an adequate physical interpretation.

As is known the equation of motion for the $F(R)$ theory of gravity written in the Friedmann-Lemaître-Robertson-Walker (FLRW) metric includes both derivatives of different orders with respect to the curvature of function $F(R)$ and derivatives with respect to time of the Hubble rate $ H $ with a differential relation between $ R $ and $ H $. This makes it difficult to solve these equations in general and many authors specify firstly the desired behaviour of the Hubble rate and then find a solution for $ F (R) $. So in the works \cite{NO_2006}-\cite{NOSG_2009} Nojiri \textit{et al} developed the scheme for cosmological reconstruction of $ F (R) $ gravity which is based on using the e-foldings number $ N $ instead of the cosmic time. Using this method, the authors constructed a number of $ F (R) $ gravity examples where various scenarios of background evolution can be realized, including the $\Lambda CDM $ cosmology, the deceleration and acceleration epoch which is equivalent to presence of phantom and non-phantom matter, oscillating Universe.

In this paper we focus on the possibility of solving the Friedmann equation exactly. We impose an ansatz on the Hubble rate that depends on an arbitrary function and satisfies the following condition: this function must allow one to obtain an exact time dependence for the main physical quantities of interest: the scalar curvature, the Hubble rate, the scale factor and the function $ F (R) $. Due to the requirement for the ansatz function their number is very restricted. An interesting fact is that the solutions that deserve attention as physical ones and were previously considered by other authors were obtained in this article as corresponding to the simplest form of the ansatz function. Then we briefly analyse the obtained solutions to see if they are interesting from a physical point of view. We also considered one of the obtained solutions as possible for describing cosmological inflation.

\section{Models under consideration and field equations}

The metric $f(R)$-gravity models without matter fields has the action 
\be
S = \frac{1}{16 \pi G} \int d^{4}x \sqrt{-\textit{g}} \, F(R),
\ee
where $ G $ is the gravitational constant, $ \textit{g} $ is the determinant of the metric tensor $ g_{\mu\nu} $, $ R = g^{\mu\nu} R_{\mu\nu} $ is the scalar curvature, or the Ricci scalar, $ R_{\mu\nu} $ is the Ricci tensor.
Variation of eq.(1) with respect to the metric gives the field equations \cite{SF_2010}
\be
F_{R}(R) R_{\mu \nu} - \frac{1}{2}F(R)g_{\mu \nu} - [ \nabla_{\mu} \nabla_{\nu} - g_{\mu \nu} \square ] F_{R}(R) =  0,
\ee
where $ F_{R}(R)\equiv \dfrac{dF(R)}{dR} $. We will consider a Friedmann–Lemaître–Robertson–Walker (FLRW) metric interval:
\be
ds^{2} = - dt^{2} + (a(t))^{2} \sum \limits_{i = 1, 2, 3} (dx^{i})^{2}
\ee
and eqs. (2) take the form
\be
18 H (\ddot{H}  +  4 H \dot{H})\, F_{R R} - 3 \,(\dot{H} + H^{2}) F_{R} + \frac{1}{2} F = 0 
\ee
\be
36 (\ddot{H}  +  4 H \dot{H})^2 F_{R R R} + 6 (\dddot{H} + 6 H \ddot{H} + 4 \dot{H}^2 + 8 H^2 \dot{H})F_{R R} - (\dot{H} + 3 H^{2}) F_{R} + \frac{1}{2} F = 0,
\ee
where eq. (4) is the Friedmann equation; it can be shown that eq. (5) is a consequence of eq. (4). The scalar curvature $ R  =  6 \, (\dot{H} + 2 H^{2})  $ where the Hubble rate $ H = \frac{\dot{a}}{a} $ and the Hubble radius $ R_{H} =  \frac{1}{\dot{a}} $ 
($ \dot{ } \equiv \frac{d}{dt}$). 
Eq. (4) contains, generally speaking, differentiation with respect to two variables: $ t $ and $ R $, where may be $ \frac{\partial R}{\partial t} \neq 0 $. Therefore, the writing this equation in terms of single variable $ t $ or $ R $ without concretization of the form of $ H(t) $ and, as a consequence, $ a(t) $ and $ R(t) $ is not possible. In this article we choose for the Hubble rate $H$ a certain ansatz depending on an arbitrary function $ \Phi(H^2) $ which will be defined below to solve exactly eq. (4). It is obvious that not any predetermined function will lead to a solution that is interesting from a physical point of view. It is also clear that even with some simple form of $ \Phi(H^2) $ eq. (4) may no longer be solved in quadratures. Also a necessary condition is the ability to obtain explicit expressions for the scalar curvature $ R(t) $, the Hubble rate $ H(t) $, the scale factor  $ a(t) $ and preferably for the function $ F (R) $. Nevertheless it seems interesting to us to obtain different solutions of Friedmann equation by choosing the simplest forms of the function $ \Phi(H^2) $ introduced in the ansatz. 

Further one can see that we can substitute our ansatz into eq. (4) and obtain solutions depending on the variable t. However for the convenience of comparing the resulting solutions with solutions obtained previously by other authors we will use the cosmological reconstruction technique developed in \cite{NO_2006}-\cite{OO_2021}. This technique allows to pass from the cosmological time variable $ t $ to a new e-folding number variable $ N = \ln{\frac{a}{a_{0}}} $
and rewrite eq. (4) by using a new variable as:
\be
9 H^{2} \big((H^{2})'' + 4 (H^{2})' \big)\, F_{R R} - 3 \,(H^{2} + \frac{1}{2}\,(H^{2})') F_{R} + \frac{1}{2} F = 0, \,\,\,\, {where \,\, ' \equiv \frac{d}{dN}}.
\ee

\section{The chosen ansatz and the investigated equation}

Let us now choose an ansatz that will simplify the solution of eq. (4): we will consider $ \frac{1}{\sqrt{H^{2}}} \frac{d H^{2}}{dt} $ as a function of $ H^{2} $. If we write this from e-foldings number variable defined in terms of the scale factor as $ e^{N} = \frac{a}{a_{0}} $ we obtain:
\be
\frac{d H^{2}}{dN} = \Phi(H^{2})
\ee
so that 
\be
R  =  3 \, \big(\Phi(H^{2}) + 4 H^{2}\big).
\ee
We introduce the notation $ H^{2} \equiv x $ and rewrite the functions entering the eq. (4) in terms of $ x $. Because
\ba
\dot{H} & = & \frac{1}{2} \Phi(x) \\ \nonumber
\ddot{H} & = & \frac{1}{2} \sqrt{x}   \, \Phi \Phi_{x}  \\ \nonumber
F_{R} & = & \frac{F_{x}}{3(\Phi_{x} + 4)}  \\ \nonumber
F_{RR} & = & \frac{ F_{xx}}{9(\Phi_{x} + 4)^{2}} - \frac{ \Phi_{xx} \, F_{x}}{9(\Phi_{x} + 4)^{3}} 
\ea
and $R  =  3 \, (\Phi(x) + 4 x) $
eq. (4) takes the form:
\be
F_{xx} - \Big(\frac{\Phi_{xx}}{\Phi_{x} + 4} + \frac{\Phi + 2 x}{2 \Phi x}\Big)\, F_{x} + \frac{\Phi_{x} + 4}{2 \Phi x} \, F = 0.
\ee
Thus we obtain the equation in which all functions depend on one variable $ x $. The chosen ansatz of course imposes a restriction on the solutions obtained for $ F $. However this allows one to specify the function $\Phi(x)$ arbitrarily and for this given $\Phi(x)$ to solve this equation exactly. Obviously not for any given function $ \Phi(x) $ the equation can be solved in known functions. As we will see below even in some of the simplest cases the solution for $ F(x) $ will be representable only as a series. Further since we are interested in the dependence of $ F $ on the curvature $ R $, it is also necessary that we can express exactly $ x(R) $ from eq. (8). And of course the determining factor in choosing of $ \Phi(x) $ is whether the resulting solution is interesting from a physical point of view. So by choosing $ \Phi(x) $ we can immediately solve eq. (7) for $ H = H(N) $:
\be
\int\frac{dx}{\Phi(x)} = N + N_{0},
\ee
where $ N_{0} = const $ and taking into account that $ H = \frac{dN}{dt} $ we obtain $ N(t) $:
\be
\int\frac{dN}{\sqrt{x(N)}} = t + t_{0}.
\ee
If we were only interested in $ a(t) $ then we could not find dependence $ N(t) $ and immediately get $ H(t)\equiv \sqrt{x} = \frac{\dot{a}}{a} $. But we find it useful to us to obtain the expression for time dependence of e-foldings number.

\section{Exact solutions}

Let us now consider several examples of obtaining solutions to the eq. (10).

\subsection{$ \Phi(x) = C_{0} $, $ C_{0} = const $}

This case corresponds to the exact solution found and analyzed in detail by Odintsov and Oikonomou in the work \cite{OO_2021}.

\subsection{$ \Phi(x) = C_{0} x + \tilde{C_{00}} $, $ C_{0} = const $, $ C_{0} \neq -4 $, $ \tilde{C_{00}} = const $} 

Substituting this $\Phi(x) $ into eq. (10) leads to the equation:
\be
2x \, (C_{0} x + \tilde{C_{00}}) F_{xx} - \big( (C_{0} + 2) x + \tilde{C_{00}} \big) F_{x} + (4 + C_{0}) F =0.
\ee
Because of eq. (8) the scalar curvature is 
\be
R = 3 \Big((C_{0} + 4) x + \tilde{C_{00}} \Big)
\ee
and one can easily pass from differentiating with respect to $ x $ to differentiating with respect to $ R $. It is obvious that the resulting solutions for $ F $ will depend on the value of the constants included in eq. (13). The function $ x(N) $ has the form:
\be
x = \tilde{C_{0}} e^{C_{0}N} - \frac{\tilde{C_{00}}}{C_{0}},\,\,\, \tilde{C_{0}} = const.
\ee
This type of function was studied in the works \cite{NO_2006}-\cite{NOSG_2009}, we will discuss this below. E-foldings number $ N(t) $ can be found from the integral:
\be
\int \frac{dN}{\sqrt{\tilde{C_{0}} e^{C_{0}N} - \frac{\tilde{C_{00}}}{C_{0}}}} = t + \tilde{t_{0}},\,\,\, \tilde{t_{0}} = const
\ee
and the value of this integral depends on the sign of the constants included in the integrand. Let us consider four cases.

\textbf{\textit{Case 1}:   \textit{$ \frac{\tilde{C_{00}}}{C_{0}} = 0 $}}

For $ \tilde{C_{00}} = 0 $ we obtain the Euler equation
\be
x^{2} \, F_{xx} - \frac{2 + C_{0}}{2 C_{0}} \, x \, F_{x} + \frac{4 + C_{0}}{2 C_{0}} \, F = 0.
\ee
whose solution has the form: 
\be
F(R) = C_{+} R^{k_{+}} + C_{-} R^{k_{-}}
\ee
for 
\ba
k_{\pm} & = & \frac{1}{4} \Bigg[ 3 + \frac{2}{C_{0}} \pm  \sqrt{1 - \frac{20}{C_{0}}  + \frac{4}{C_{0}^{2}}}\Bigg],\\ \nonumber 
C_{0} & \in &(- \infty;0) \cup ( 0; 2( 5-2\sqrt{6})) \cup (2(5+2\sqrt{6}; \infty  ); 
\ea
\be
F(R) = C_{1} R^{\frac{4 \pm \sqrt{6}}{2}} + C_{2} R^{\frac{4 \pm \sqrt{6}}{2}} \ln R, \,\, 
\ee
for $ C_{0}  = 2( 5 \pm 2\sqrt{6}) $
and 
\be
F(R) = C_{3} R^{\frac{1}{4} ( 3 + \frac{2}{C_{0}})} \, \cos \big( \frac{\sqrt{|C_{0}^{2} - 20 C_{0} +4|}}{|C_{0}|} \, \ln R  + \varphi_{0} \big)
\ee
for $C_{0}  \in (2( 5-2\sqrt{6}); 2(5+2\sqrt{6}) ) $; $ C_{i} $, $ i = 1, 2, 3, \pm $ and $ \varphi_{0} $ are constants. 

The scale factor $ a(t) $ is 
\ba
a(t) & = & a_{0} \, (\frac{-C_{0} \sqrt{\tilde{C_{0}}} }{2} \, t + t_{0})^{-\frac{2}{C_{0}}}, 
\ea
the function $ x(t) $ and e-foldings number $ N(t) $ are:
\ba
x(t) & = & \tilde{C_{0}} \,( \frac{-C_{0} \sqrt{\tilde{C_{0}}} }{2}  \,t + t_{0})^{-2}   \\ \nonumber
N(t) & = &  \frac{2}{|C_{0}|}\, \ln \Big(  \frac{-C_{0} \sqrt{\tilde{C_{0}}} }{2}  \, t + t_{0}\Big),
\ea
where $ \tilde{C_{0}} $ is a positive constant. Then for the scalar curvature we obtain the following expression:
\be
R(t) = \frac{3\tilde{C_{0}}(C_{0} + 4) }{( \frac{-C_{0} \sqrt{\tilde{C_{0}}} }{2}  \,t + t_{0})^{2} }.
\ee
Let us analyze the solutions obtained for $ F(R) $. It is easy to see from eqs. (18)-(19) that it cannot be simultaneously $ k_{+} \sim 1$ and $ k_{-} \sim 2$ or vice versa: $  k_{+} \in (1; 2) $, $ k_{-} \in (-\infty;  \frac{1}{2}) $ for $ C_{0} < 0 $, $  k_{+} \in [2 + \frac{\sqrt{6}}{2}; \infty) $, $ k_{-} \in (2; 2 + \frac{\sqrt{6}}{2}] $ for $ C_{0} \in (0; 2(5 - 2\sqrt{6})] $ and $  k_{+} \in [ 2 - \frac{\sqrt{6}}{2}; 1) $, $ k_{-} \in (\frac{1}{2}; 2 - \frac{\sqrt{6}}{2}] $ for $ C_{0} \geq 2(5 + 2\sqrt{6})) $. It is clear that $k_{+} \neq 1  $ and $k_{-} \neq \frac{1}{2}  $. At $ k_{+} = \frac{5}{4} $ and $ k_{-} = 0  $ we get $ C_{0} = - 4$ which cannot be according to the condition, see $ \textbf{B} $. The cases $ k = \frac{1}{2}, \frac{5}{4} $ and $ 2 $ will be considered below. 

The solution in the form of eq. (18) was obtained in work \cite{NOSG_2009} and the authors concluded that the phantom dark energy cosmology, see eq. (22) for the scale factor $ a(t) $, can be also obtained in the $ F(R) $ gravity frame without introducing a phantom fluid. Indeed, consider equation (22). For $ C_{0} < 0 $ and for $ t_{0} \geq 0 $ the scale factor is $ a(t) \sim (t + t_{0})^{\frac{2}{|C_{0}|}} $ and it is clear that $ a(t) $ grows monotonically with increasing time. Similarly the Hubble rate $  H(t) \sim (t + t_{0})^{-1} $ and the scalar curvature $ R \sim (t + t_{0})^{-2} $ are continuous functions for all $ t > 0 $. For $ t_{0} < 0 $ the domain of definition of $ a(t) $ for arbitrary negative $ C_{0} $ is $ t \geq | t_{0}|$ and for $ H $ and $ R $ this is $ t > | t_{0}|$ because at $ t = | t_{0}| $ functions $ H $ and $ R $ diverge. However, this singularity "in the past" may be purely coordinate one. 

But for positive $ C_{0}$ the functions $ a(t) \sim ( - t + t_{s})^{-\frac{2}{C_{0}}} $, $  H(t) \sim h_{0} ( -t + t_{s})^{-1}$ and $ R \sim 6 h_{0} (1 + 2 h_{0}) ( - t + t_{s})^{-2} $ are defined only for $ t < t_{s} $ and have the future singularities at $ t = t_{s} $. Since it is easy to show that $ \kappa ^{2}\rho_{eff} = 3 H^{2} \longrightarrow \infty $ and $ \kappa ^{2} | p_{eff}| = |2 \dot{H} +  3 H^{2}| \longrightarrow \infty  $ ($ \kappa^{2} = 8 \pi G_{N} $, $  G_{N}$ is the Newton’s gravitational constant) at $ t \longrightarrow t_{s} $ this is the Type I (Big Rip) singularity, see \cite{NOO}, \cite{HNOOP} (and references therein) and $ t_{s} $ is the so-called Rip time. This solution describes the universe that ends at the Big Rip singularity in the time $ t_{s} $. The equation of state parameter $ w $ is expressed in terms of $ C_{0} $ as follows: $ w = - 1 - \frac{C_{0}}{6} $ that is $ w < - 1 $ as it should be for a phantom fluid in the standard General Relativity.

\textbf{\textit{Case 2}:   \textit{$ \frac{\tilde{C_{00}}}{C_{0}} \neq 0 $}}

If we want to obtain a hypergeometric equation we can pass to the variable $ z = \frac{-C_{0} (R - 3 \tilde{C_{00}})}{3 \tilde{C_{00}} (C_{0} + 4)}$, then $F = C \,\, {_{2}F_{1}} (\alpha; \beta; -\frac{1}{2}; z) + \tilde{C} \,\, z^{\frac{3}{2}} \,\, {_{2}F_{1}} (\alpha + \frac{3}{2}; \beta + \frac{3}{2}; \frac{5}{2}; z) $ for $ z < 1 $ or $ F = C_{1} \, \, (-z)^{-\alpha} \, \, {_{2}F_{1}} (\alpha; \alpha + \frac{3}{2}; \alpha - \beta + 1; \frac{1}{z}) + \tilde{C_{1}} \,\, (-z)^{-\beta} \,\, {_{2}F_{1}} (\beta; \beta + \frac{3}{2}; \beta - \alpha + 1; \frac{1}{z}) $ for large values of $ z $ where $ \alpha = - k_{\pm} $, $ \beta = - k_{\mp} $ for $ C_{0} > 0 $ and vice versa for $ C_{0} < 0 $ and $ k_{\pm} $ are defined in eq. (19). 

Exactly the same Gauss hypergeometric function for $z < 1$ was obtained in the work \cite{NOSG_2009} for $ C_{0} = -3 $, see eq. (15). The authors consider the $ F (R) $ gravity which reproduces the $ \Lambda CDM $-era without real matter 
and thus they demonstrate that the modified gravity may describe this epoch "without the need to introduce the effective cosmological constant". 

\textbf{\textit{Case 2a}:   \textit{$ \frac{\tilde{C_{00}}}{C_{0}} > 0 $, $ \tilde{C_{0}} > 0 $}}

In this case the functions $ x(t) $, the scale factor $ a(t) $ and e-foldings number $ N(t) $ have the form:
\ba
x(t) & = & \frac{\tilde{C_{00}}}{C_{0}}\, \tan^{2} \big(\frac{\sqrt{\tilde{C_{00}}C_{0}}}{2} \, t + t_{0}\big)  \\ \nonumber
a(t) & = & a_{0} \Big| \,\sqrt{ \frac{C_{0}\tilde{C_{0}} }{\tilde{C_{00}}}} \, \cos \big(\frac{\sqrt{\tilde{C_{00}}C_{0}}}{2} \, t + t_{0}\big)\Big|^{-\frac{2}{C_{0}}}\\ \nonumber
N(t) & = &  -\frac{2}{C_{0}}\ln \Big|\sqrt{ \frac{C_{0}\tilde{C_{0}} }{\tilde{C_{00}}}} \cos \big(\frac{\sqrt{\tilde{C_{00}}C_{0}}}{2} \, t + t_{0}\big)  \Big|,
\ea
$ t_{0} = const $. The scalar curvature $ R $ is:
\be
R(t) = 3 \tilde{C_{00}} \big(1 + \frac{C_{0} + 4}{C_{0}} \tan^{2} \big(\frac{\sqrt{\tilde{C_{00}}C_{0}}}{2} \, t + t_{0}\big) \big)
\ee
This case seems unrealistic to us since the periodic behaviour of all functions and cannot have a reasonable interpretation.

\textbf{\textit{Case 2b}:   \textit{$ \frac{\tilde{C_{00}}}{C_{0}} < 0 $, $ \tilde{C_{0}} > 0 $}}

For these values of the constants included in eq. (13) we obtain the following functions for $ x(t) $, the scale factor $ a(t) $ and e-foldings number $ N(t) $:
\ba
x(t) & = & \Big|\frac{\tilde{C_{00}}}{C_{0}}\Big| \, \coth^{2} \big(
\frac{C_{0}}{2} \sqrt{\Big| \frac{\tilde{C_{00}}}{C_{0}}\Big|} \, t + t_{0}\big) \\ \nonumber
a(t) & = & a_{0} \, \Big|\sqrt{\Big| \frac{C_{0}\tilde{C_{0}} }{\tilde{C_{00}}}\Big|} \,\sinh \big(
\frac{C_{0}}{2} \sqrt{\Big| \frac{\tilde{C_{00}}}{C_{0}}\Big|} \, t + t_{0}\big) \Big|^{-\frac{2}{C_{0}}}\\ \nonumber
N(t) & = & - \frac{2}{C_{0}}\ln \Big|\sqrt{\Big| \frac{C_{0}\tilde{C_{0}} }{\tilde{C_{00}}}\Big|} \sinh \big(\frac{C_{0}}{2} \sqrt{\Big| \frac{\tilde{C_{00}}}{C_{0}}\Big|} \,t + t_{0}\big)  \Big|,
\ea
$ t_{0} = const $. If we put $ C_{0} = -3 $, $ t_{0} = 0 $ and $ \sqrt{\Big| \frac{\tilde{C_{00}}}{C_{0}}\Big|} =  H_{0} \sqrt{\Omega_{vac}} $ in expression for the scale factor, we obtain the general GR solution for a flat universe with matter and vacuum energy \cite{FTH_2008} ($ \Omega_{vac} $ is the normalizing vacuum energy density, $ H_{0} $ is the present value of the Hubble parameter). An expression for the scale factor as a power of the hyperbolic sine was also obtained in \cite{NO_2006}. The expression for scalar curvature which reproduces the scale factor of flat space has the form:
\be
R(t) = 3  \big(\tilde{C_{00}} + \frac{(C_{0} + 4)|\tilde{C_{00}}| }{|C_{0}|} \coth^{2} \big(
\frac{C_{0}}{2} \sqrt{\Big| \frac{\tilde{C_{00}}}{C_{0}}\Big|} \, t + t_{0}\big)\big).
\ee

\textbf{\textit{Case 2c}:   \textit{$ \frac{\tilde{C_{00}}}{C_{0}} < 0 $, $ \tilde{C_{0}} < 0 $}}

For the fourth case of choosing coefficient correspond the following functions for $ x(t) $, the scale factor $ a(t) $, e-foldings number $ N(t) $ and the curvature $ R $:
\ba
x(t) & = & \Big|\frac{\tilde{C_{00}}}{C_{0}}\Big| \, \tanh^{2}\big(
\frac{C_{0}}{2} \sqrt{\Big| \frac{\tilde{C_{00}}}{C_{0}}\Big|} \, t + t_{0}\big)   \\ \nonumber
a(t) & = & a_{0} \, \Big(\cosh \big(
\frac{C_{0}}{2} \sqrt{\Big| \frac{\tilde{C_{00}}}{C_{0}}\Big|} \, t + t_{0}\big) \Big)^{-\frac{2}{C_{0}}}\\ \nonumber
N(t) & = & - \frac{2}{C_{0}}\ln \Big(\sqrt{\Big| \frac{C_{0}\tilde{C_{0}} }{\tilde{C_{00}}}\Big|} \cosh \big(\frac{C_{0}}{2} \sqrt{\Big| \frac{\tilde{C_{00}}}{C_{0}}\Big|} \,t + t_{0}\big)  \Big)\\ \nonumber
R(t) & = & 3  \big(\tilde{C_{00}} + \frac{(C_{0} + 4)|\tilde{C_{00}}| }{|C_{0}|} \tanh^{2} \big(
\frac{C_{0}}{2} \sqrt{\Big| \frac{\tilde{C_{00}}}{C_{0}}\Big|} \, t + t_{0}\big)\big),
\ea
$ t_{0} = const $.
It can be seen that for $ C_{0} = 1 $ the solution for $ a(t) $ formally has the form of a Korteweg-de Vries soliton. However the KdV-soliton is a travelling wave one but eq. (29) describes only a time dependence. Therefore we assume that this similarity does not hide a deep meaning.

\subsection{$ \Phi(x) = \alpha   x^{2} $} 

It is clear that not every simple function $ \Phi(x) $ even allows one to formally obtain expressions for $ N(t) $ or $ a(t) $. So at $ \Phi(x) = \alpha x^{2} + \beta $, $ \beta \neq 0 $ the integral in eq. (12) is taken exactly but one cannot to express $ N = N(t) $. Consider the case of $ \beta = 0 $. The function $ x(N) $ has the form:
\be
x = -\frac{1}{\alpha (N + N_{0} )}
\ee
that gives $ N(t) $, $ a(t) $ and $ R(t) $ as follow:
\ba
N(t) & = & - \frac{1}{\alpha} \, \big(\frac{- 3 \alpha}{2}  t  + t_{0} \big)^{\frac{2}{3}} - N_{0} \nonumber \\
a(t) & = & a_{0} e^{ - \frac{1}{\alpha} \, \big(\frac{ -3 \alpha}{2}  t  + t_{0} \big)^{\frac{2}{3}} - \, N_{0} }\nonumber \\
R(t) & = & \frac{3}{\big(\frac{ -3 \alpha}{2}  t  + t_{0} \big)^{\frac{2}{3}}} \,\,\bigg( 4 +\frac{\alpha}{\big(\frac{- 3 \alpha}{2}  t  + t_{0} \big)^{\frac{2}{3}}} \bigg ),
\ea
$ N_{0} $ and $ t_{0} $ are constants. For $ \alpha > 0 $ and positive $ t_{0} $ the scale factor grows exponentially until $ t = t_{s} $; the Hubble rate $ H \sim (t_{s} - t)^{-\frac{1}{3}} $ and the curvature $ R \sim (t_{s} - t)^{-\frac{4}{3}} $ are also defined for $ t < t_{s} $ and diverge for $ t = t_{s} $. Since $ \kappa ^{2}\rho_{eff} \longrightarrow \infty $ and $ \kappa ^{2} | p_{eff}| \longrightarrow \infty  $ at $ t \longrightarrow t_{s} $  this may correspond to a Type III (Big Freeze) singularity \cite{NOO}, \cite{HNOOP}. For $ \alpha < 0 $ the scale factor grows exponentially over the entire domain of definition. Further eq. (10) has the form:
\be
2 \alpha x^{3} \, (4 + 2 \alpha x ) F_{xx} - \big( 6 \alpha^{2} x^{3} + 8 \alpha x^{2} + 8 x \big) F_{x} + (4 + 2 \alpha x)^{2} F =0.
\ee
For $ \alpha = - \frac{1}{2} $ we find the exact solution of this equation explicitly:
\be
F(x) = C_{1} \sqrt{x} \, e^{\frac{2}{x}} + C_{2} \Big( \frac{2}{3} x^{2} - \frac{32}{3} x - \frac{32\sqrt{2\pi}}{3} \, \sqrt{x} \, e^{\frac{2}{x}} \,\, erf\big(\sqrt{\frac{2}{x}}\big) \Big) 
\ee
where $ x $ is expressed in terms of $ R $ as follows:
\be
x = 4 \pm \sqrt{16 - \frac{2}{3} R}.
\ee

\subsection{$ \Phi(x) = \alpha \sqrt{x} $, $ \alpha > 0 $} 

One of the simplest forms of the function $  \Phi(x) $ is  $ \Phi(x) \sim \sqrt{x} $, however already in this case we find solutions of eq. (10) only in the form of series. The function $ x(N) $ has the form:
\be
x = \frac{\alpha^{2}}{4} (  N + N_{0} )^{2}
\ee
that gives $ N(t) $ and $ a(t) $ as
\ba
N(t) & = & C_{0} \, e^{\frac{\alpha}{2} t } - N_{0} \nonumber \\
a(t) & = & a_{0} \, e^{ C_{0} \,\, e^{\frac{\alpha}{2} t } - N_{0}}
\ea
where $ N_{0} $ and $ C_{0} $ are constants. The scalar curvature is:
\be
R(t) = \frac{3 \alpha^{2} C_{0}}{2} \,\, e^{\frac{\alpha}{2} t } \, \big( 1 + 2 C_{0} e^{\frac{\alpha}{2} t } \big).
\ee
If we denote $ \alpha = - 2 \lambda $, $ N_{0} = - \ln C_{1} $ and express $ N $ in terms of $ R $ we obtain exactly the solution found in the work \cite{O_2018}. Eq. (10) has the form:
\be
\alpha x^{2} \, (\alpha + 8 \sqrt{x}  ) F_{xx} - x\sqrt{x} \big( 5 \alpha + 8 \sqrt{x} ) F_{x} + \frac{1}{4}(\alpha + 8 \sqrt{x} )^{2} F =0.
\ee
For the convenience of finding a solution let us move on to the new variable $  \tilde{x} = \sqrt{x}$ and to the new function $ \tilde{F} = \tilde{x}^{-1}  F $ and write eq. (38) as:
\be
\tilde{F}_{\tilde{x}\tilde{x}} + \frac{-\frac{2}{\alpha} \tilde{x}^{2} - \frac{1}{4} \tilde{x} + \frac{\alpha}{8}   }{\tilde{x} \, (\tilde{x} + \frac{\alpha}{8} )} \,\, \tilde{F}_{\tilde{x}} + \frac{\frac{6}{\alpha} \tilde{x} - \frac{1}{4}}{ \tilde{x} \, (\tilde{x} + \frac{\alpha}{8} )} \,\, \tilde{F} =0.
\ee
Expanding in a Laurent series the coefficients at $  \tilde{F}_{\tilde{x}} $ and $ \tilde{F} $ as 
\ba
\frac{-\frac{2}{\alpha} \tilde{x}^{2} - \frac{1}{4} \tilde{x} + \frac{\alpha}{8}   }{\tilde{x} \, (\tilde{x} + \frac{\alpha}{8} )} & = & \frac{1}{\tilde{x}} - \frac{10}{\alpha} - \frac{8}{\alpha} \, \sum\limits_{n=1}^\infty \, (- \frac{8}{\alpha})^{n} \, \tilde{x}^{n} \nonumber \\
\frac{\frac{6}{\alpha} \tilde{x} - \frac{1}{4}}{ \tilde{x} \, (\tilde{x} + \frac{\alpha}{8} )} & = & -\frac{2}{\alpha \tilde{x}} + \frac{64}{\alpha^{2}} + \frac{64}{\alpha^{2}} \, \sum\limits_{n=1}^\infty \,(- \frac{8}{\alpha})^{n} \, \tilde{x}^{n}
\ea
we are looking for two linearly independent solutions of the eq. (39) in the form of series \cite{ZP}. Passing to the initial variable $ x $ and function $ F $ we obtain: 
\ba
F_{1}(x) & = & \sqrt{x} \, \Big(  1 + \sum\limits_{n=1}^\infty \, a_{n} x^{\frac{n}{2}} \Big) \nonumber \\
F_{2}(x) & = & F_{1}(x) \, \ln \sqrt{x} + \sum\limits_{n=1}^\infty \, b_{n} x^{\frac{n +1}{2}}.
\ea
From eq. (8) for $ R$ we obtain for positive $ H $ the expression:
\be
 H  =  \frac{1}{8} \big( - \alpha + \sqrt{\alpha ^{2} + \frac{16R}{3}} \big) 
\ee
and substituting this into eq. (41) we obtain expressions for $ F(R) $. The values of the coefficients $ a_{i} $ and $ b_{i} $ in the series are found by substituting eqs. (41) into eq. (38): 
\ba
a_{1} & = & \frac{2}{\alpha} \nonumber \\
a_{2} & = & - \frac{10}{\alpha^{2}} \nonumber \\
a_{3} & = & \frac{292}{9\alpha^{3}}\nonumber \\
... \nonumber \\
a_{p+2} & = & -\frac{1}{(p+2)^{2}} \, \Big( - \frac{2(5p+6)}{\alpha} a_{p+1} + \frac{64}{\alpha^{2}} \big(a_{p} + (-\frac{8}{\alpha})^{p} \big) + \nonumber \\ & + & \sum\limits_{k=1}^p \, (-\frac{8}{\alpha})^{k+1} \, (p - k +1) a_{p-k+1} + \frac{64}{\alpha^{2}} \, \sum\limits_{k=1}^{p-1} \, (-\frac{8}{\alpha})^{k} \,  a_{p-k} \Big); \,\,\, p \geqslant 2
\nonumber \\
\ea
where the value of the next coefficient is recursively expressed in terms of the previous ones. For coefficients $ b_{i} $ the procedure is similar.

\subsection{$ \Phi(x) = \alpha \sqrt{x + \sigma |\beta|} $, $ \alpha > 0 $, $ \sigma = \pm 1 $} 

A slight complication of the previous function $  \Phi(x) $ does not lead to significant changes. The function $ x(N) $ has the form:
\be
x = \frac{\alpha^{2}}{4} (  N + N_{0} )^{2} - \beta. 
\ee
The folding number $ N(t)$ is
\be
N(t)  =  \begin{cases}
\sigma = + 1 :  &  \frac{2\sqrt{|\beta |}}{\alpha} \cosh \big(\frac{\alpha}{2} (t - t_{0}) - \frac{1}{2} \ln |\beta | \big) - N_{0} \\
\sigma = - 1:   &  \frac{2\sqrt{|\beta |}}{\alpha} \sinh \big(\frac{\alpha}{2} (t - t_{0}) - \frac{1}{2} \ln |\beta | \big) - N_{0}  
\end{cases},
\ee
$ N_{0} $ and $ t_{0} $ are constants, $ \frac{\alpha}{2} (t - t_{0}) - \frac{1}{2} \ln |\beta | > 0 $, that gives 
\be
x(t)  =  \begin{cases}
\sigma = + 1 :  &  |\beta | \sinh^{2} \big(\frac{\alpha}{2} (t - t_{0}) - \frac{1}{2} \ln |\beta | \big) \\
\sigma = - 1:   &  |\beta | \cosh^{2} \big(\frac{\alpha}{2} (t - t_{0}) - \frac{1}{2} \ln |\beta | \big)
\end{cases}
\ee
and the scale factor has the form:
\be
a(t)  = a_{0} \exp \Big( \frac{2 \sqrt{|\beta|}}{\alpha} \, \begin{cases}
\sigma = + 1:  &   \cosh \big(\frac{\alpha}{2} (t - t_{0}) - \frac{1}{2} \ln |\beta | \big) - N_{0} \\
\sigma = - 1:   &  \sinh \big(\frac{\alpha}{2} (t - t_{0}) - \frac{1}{2} \ln |\beta | \big) - N_{0}
\end{cases} \Big).
\ee
The scalar curvature is:
\be
R(t) = 3 \sqrt{|\beta |}  \begin{cases}
\sigma = + 1 :  &  \cosh \big(\frac{\alpha}{2} (t - t_{0}) - \frac{1}{2} \ln |\beta | \big) \cdot \Big( \alpha +  4 \sqrt{|\beta|} \cosh \big(\frac{\alpha}{2} (t - t_{0}) - \frac{1}{2} \ln |\beta | \big) \Big) \\
\sigma = - 1:   &  \sinh \big(\frac{\alpha}{2} (t - t_{0}) - \frac{1}{2} \ln |\beta | \big) \cdot \Big( \alpha +  4 \sqrt{|\beta|} \sinh \big(\frac{\alpha}{2} (t - t_{0}) - \frac{1}{2} \ln |\beta | \big) \Big)
\end{cases}.
\ee
Eq. (10) in terms of the variable $  \tilde{x} = \sqrt{x + \sigma |\beta|}$ has the form:
\be
F_{\tilde{x}\tilde{x}} + \frac{-4 \tilde{x}^{3} -\frac{9 \alpha}{2} \tilde{x}^{2} - (\frac{\alpha^2}{4} - 4 \sigma |\beta|)  \tilde{x} + \frac{\alpha \sigma |\beta|}{2}   }{( \tilde{x}^{2} - \sigma |\beta|) \, (8 \tilde{x} + \alpha )} \,\, F_{\tilde{x}} + \frac{(4 \tilde{x} + \frac{\alpha}{2})^2}{ ( \tilde{x}^{2} - \sigma |\beta|) \, (8 \tilde{x} + \alpha )} \,\,  F  =0.
\ee
Changing to a new function $ \tilde{F} $ (see eq. (39)) does not simplify this equation and the solution can again be found only in the form of a series. The procedure for constructing a series is similar to the previous case.

\subsection{$ \Phi(x) = \frac{\alpha}{x} $} 

Let us consider another example of a fairly simple function for which it is impossible to obtain an explicit expression for $ N(t) $. So for $ \Phi(x) = \frac{\alpha}{x} + \beta + \gamma x $ the integral (12) can be easily taken for arbitrary values of the constants $ \alpha $, $ \beta $ and $\gamma  $. But this results in a differential equation for $ N(t) $ which is solved parametrically, that does not allow to obtain an easily interpretable expression for $ a(t) $. By introducing some relations for the constants it is possible to simplify the integral (12) but it will still not be possible to express $ N(t) $. Consider now the simplest case corresponding to $  \beta = \gamma = 0 $. The function $ x(N) $ has the form:
\be
x = \sqrt{2 \alpha \, ( N + N_{0} )}
\ee
and e-folding number is
\be
2 \alpha \, ( N + N_{0} ) =  \big( \frac{3 \alpha}{2} t + t_{0} \big)^{\frac{4}{3}}.   \nonumber \\
\ee
The scale factor end the scalar curvature are:
\ba
a(t) & = & a_{0} \, e^{\frac{1}{2 \alpha} \, (\frac{3 \alpha}{2} t + t_{0})^{\frac{4}{3}} - N_{0}} \nonumber \\
R(t) & = & \frac{\alpha}{(\frac{3 \alpha}{2} t + t_{0})^{\frac{2}{3}} } + 4 (\frac{3 \alpha}{2} t + t_{0})^{\frac{2}{3}},
\ea
where $ N_{0} $ and $ t_{0} $ are constants. Consider $ \alpha < 0 $ and $ t_{0} > 0$. The scale factor $ a \leq a_{s} $ at $ t \leq t_{s} $. Despite the fact that $ H \sim (t_{s} - t)^{\frac{1}{3}} $ and it does not have a singularity the curvature is singular at $ t = t_{s} $: $ R \sim (t_{s} - t)^{- \frac{2}{3}} $. Since $\kappa ^{2}\rho_{eff} = 0$ and $ \kappa ^{2} | p_{eff}| \longrightarrow \infty  $ at $ t \longrightarrow t_{s}$ this case can correspond to the finite-time
singularity of Type II (Sudden) \cite{NOO}, \cite{HNOOP}. Then eq. (10) has the form:
\be
 F_{xx} - \frac{ 4 x^{4} + \alpha x^{2} + \frac{3\alpha^{2}}{2} }{\alpha x \, (4 x^{2} - \alpha )} F_{x} + \frac{4 x^{2} - \alpha}{2 \alpha x^{2}} \, F =0.
\ee
As in the previous paragraph we will look for two linearly independent solutions of this equation in the form of series \cite{ZP}. Expanding in a Laurent series the coefficients at $  F_{x} $ and $ F $ as  
\ba
- \frac{ 4 x^{4} + \alpha x^{2} + \frac{3\alpha^{2}}{2} }{\alpha x \, (4 x^{2} - \alpha )} & = & \frac{3}{2x} + \frac{7}{\alpha} x  + \frac{32}{\alpha^{2}} x^{3} + \sum\limits_{n=5}^\infty \, \frac{2^{n+2}}{\alpha^{\frac{n+1}{2}}} \, x^{n} \\ \nonumber
\frac{4 x^{2} - \alpha}{2 \alpha x^{2}}  & = &  -\frac{1}{2x^{2}} + \frac{2}{\alpha}.
\ea
Substituting this into eq. (52) we obtain expressions for $ F(R) $:
\ba
F_{1} & = & \sqrt{x} \, \Big(  1 + \sum\limits_{n=1}^\infty \, a_{n} x^{n} \Big) \nonumber \\
F_{2} & = & \frac{1}{x} \, \Big(  1 + \sum\limits_{n=1}^\infty \, b_{n} x^{n }\Big).
\ea
The values of the coefficients $ a_{i} $ and $ b_{i} $ are found by substituting eqs. (54) into eq. (52): 
\ba
a_{1} & = & 0 \nonumber \\
a_{2} & = & - \frac{11}{14 \alpha} \nonumber \\
a_{3} & = & 0 \nonumber \\
a_{4} & = & - \frac{19}{616\alpha^{2}}\nonumber \\
a_{5} & = & 0 \nonumber \\
a_{6} & = & - \frac{3}{1232 \alpha^{3}} \nonumber \\
... \nonumber \\
a_{p+2} & = & -\frac{1}{p^{2} + 5.5 p +7} \,  \Big(\big( \frac{7p}{\alpha} + \frac{11}{2 \alpha} \big) a_{p} + (\frac{32p}{\alpha^{2}} - \frac{48}{\alpha^{2}})a_{p-2}  + \nonumber \\ & + & \big( \frac{4}{\alpha} \big)^{\frac{p+2}{2}} \, + \sum\limits_{k=1}^{p-4} \, (k + \frac{1}{2}) \,  a_{k} \, \sum\limits_{m=1}^{\frac{p-2-n}{2}} \, \frac{2^{2m+5}}{\alpha^{m+2}} \Big); \,\,\, p \geqslant 5
\nonumber \\
\ea
where the value of the next coefficient is recursively expressed in terms of the previous ones. As one can see $ a_{2q+1} = 0 $. For coefficients $ b_{i} $ the similar procedure gives:
\ba
b_{1} & = & 0 \nonumber \\
b_{2} & = &  \frac{5}{\alpha} \nonumber \\
b_{3} & = & 0 \nonumber \\
b_{4} & = & - \frac{13}{10\alpha^{2}}\nonumber \\
b_{5} & = & 0 \nonumber \\
b_{6} & = &  \frac{89}{9 \alpha^{3}} \nonumber \\
... \nonumber \\
b_{p+2} & = & -\frac{1}{p^{2} + 2.5 p +1} \,  \Big(\big( \frac{7p}{\alpha} - \frac{5}{\alpha} \big) b_{p} + (\frac{32p}{\alpha^{2}} - \frac{96}{\alpha^{2}})b_{p-2}  - \nonumber \\ 
& - &  \frac{2^{p+3}}{\alpha^{\frac{p+2}{2}}} \, + \sum\limits_{k=0}^{p-5} \, 
k b_{k} \, \sum\limits_{m=1}^{\frac{p-3-k}{2}} \, \frac{2^{2m+5}}{\alpha^{m+2}} \Big); \,\,\, p \geqslant 5 \nonumber \\.
\ea
One can also see that $ b_{2q+1} = 0 $. Expressing $ x $ in terms of $ R $ we obtain
\be
x  = \frac{1}{8} \Big( \frac{R}{3} \pm \sqrt{\frac{R^{2}}{9} - 16 \alpha} \Big)
\ee
and substituting this into eq. (54) one obtain the function $ F(R) $.

Now let us do the opposite - set the function type $F(R)$ and solve the eq. (6) exactly.

\section{$F(R)\sim R^k$, $ k \neq 0; 1 $}

Let us return to eq. (6) and substitute into it the function $F(R)\sim R^k$. Then 
\be
-9k(k-1) H^2 \Bigg((H^2)' + 4 H^2\Bigg)' + 3kR(H^2 +\frac{1}{2} (H^2)') - \frac{1}{2}R^{2} = 0
\ee
and given that  $ R = 3((H^2)' + 4 H^2) $ we obtain
\be
-3k(k-1) H^2 R' + \frac{k}{2} R(R - 6 H^2 ) - \frac{1}{2}R^{2} = 0
\ee
Denote $A \equiv -3k(k-1) $, $ B \equiv \frac{3(k-1)}{2} $. Then
\be
A H^2 \Bigg[ R \, e^{-\frac{3k}{A} N}  \Bigg]' \, e^{\frac{3k}{A} N} + B \, R ((H^2)' + 4 H^2) =0 
\ee
or
\be
R^{A} \, H^{2B} e^{(-3k+4B) N} = A_{0}, 
\ee
where $ A_{0} $ is the integration constant and $-3k+4B = 3(k-2)$. The case of $ A_{0} = 0 $ gives $ R = 0 $ and $ a = a_{0} (t+t_{0})^{\frac{1}{2}} $. It can be shown that for $ A_{0} \neq 0 $ we obtain eqs. of \textit{Case 1} again for $ k \neq 0; \frac{1}{2}; 1; \frac{5}{4}; 2 $. Thus one need to consider these values of $ k $ separately.   

For $ k = \frac{1}{2} $ we obtain $ H \sim e^{-2 N + \frac{A_{0}^{\frac{4}{3}}}{12}\, e^{6N}}$, but for the scale factor $ a $ we get the integral $ t \sim \int a e^{ -\frac{A_{0}^{\frac{4}{3}}}{12} a^{6}} d a$ which does not allow to express $ a $ explicitly as a function of $ t $. For $ k = \frac{5}{4} $ the Hubble rate is $ H \sim e^{-2 N} (N + N_{0})^{\frac{5}{6}}$, and similarly to the previous case the integral $ t \sim \int a (\ln a + \tilde{a})^{-\frac{5}{6}} d a$ is not taken in elementary functions. In case $ n = 2 $ we have $ H \sim (e^{-3 N} + N_{0})^{\frac{2}{3}}$, $ N_{0} \sim A_{0} \neq 0 $ and                                                                                                                                                                                                                                                                                                                                                                                                                                                                                                                                                                                                                                                                                                                                                                                                                                                                                                                                                                                                                                                                                                                                                                                                                                                                                                                                                                                                                                                                                                                                                                                                                                                                                                                                                                                                                                                                                                                                                                                                                                                                                                                                                                                                                                                                                                                                                                                                                                                                                                                                                                                                                                                                                                                                                                                                                                                                                                                                                                                                                                                                                                                                                                                                                                                                                                                                                                                                                                                                                                                                                                                                                       
$ t \sim \int a ( a^{3} +\tilde{N_{0}})^{-\frac{2}{3}} d a $, but despite the fact that this integral is taken exactly in elementary functions, it is impossible to express the function $ a = a(t)  $.

\section{Cosmological interpretation}

Let us now consider whether some of the solutions obtained can be related to physical solutions in cosmology and discuss their cosmological applications. We will analyse the inflationary phenomenology for $\textbf{\textit{Case 2c}}$ assuming the slow-roll dynamics. Let us look at the slow-roll indices and examine whether they can be much less than one. From eqs. (29) one may see that the Hubble rate is 
\be
H = - \sqrt{\Big|\frac{\tilde{C_{00}}}{C_{0}}\Big|} \, \tanh\big(
\frac{C_{0}}{2} \sqrt{\Big| \frac{\tilde{C_{00}}}{C_{0}}\Big|} \, t + t_{0} \big)
\ee
whence it immediately follows that 
\be
\frac{C_{0}}{2} \sqrt{\Big| \frac{\tilde{C_{00}}}{C_{0}}\Big|} \, t + t_{0} < 0.
\ee
The expression $ \frac{\ddot{H}}{H \dot{H} } \ll 1 $ gives $ C_{0} \ll 1 $ that is it is not fulfilled identically but imposes the condition on the constant $ C_{0} $. Next we will consider $ 0 < C_{0} \ll 1  $ and from eq. (63) one can see that $ t_{0} = -|t_{0}| < 0 $. Thus in this model we have a time limit $ t < t_{cr} $ where we will call $ t_{cr} = \frac{2 |t_{0}|}{\sqrt{C_{0} |\tilde{C_{00}}|}} $ as a critical time. One can see from expression for $ t_{cr}  $ that this moment in time can be made arbitrarily large.

The equation for e-folding number $ N = \int\limits_{t_{i}}^{t_{f}} Hdt $ with respect to $ t_{i} $ allows to find the time instance at the first horizon crossing (\cite{OO_2021} and references therein). The upper limit of the integral $ t_{f} < t_{cr} $ is the time of end of inflation, we will define it below. For $  t_{i} <  t_{f} $ we obtain
\be
t_{i} = \frac{2 \Big(  |t_{0}| + \ln\big[ e^{\frac{C_{0}N}{2}} \sqrt{1 + \frac{C_{0}} {2}}  - 
\sqrt{ e^{C_{0}N}  (1 + \frac{C_{0}} {2})  -  1  } \big]  \Big)}{\sqrt{C_{0} |\tilde{C_{00}}|}}.
\ee
Further we consider whether the slow-roll indices $ \epsilon_{1} $, $ \epsilon_{3} $ and $\epsilon_{4} $ can be very small \cite{NOO}, \cite{OO_2021}, \cite{HN}. The condition $ \epsilon_{1} = - \frac{\dot{H}}{H^{2} } \ll 1 $ leads to 
\be
\frac{C_{0}}{2 \sinh^{2}\big(
\frac{\sqrt{C_{0} |\tilde{C_{00}}|}}{2} \, t_{i} - |t_{0}| \big) } \ll 1 
\ee
that is $ \sinh^{2}\big(\frac{\sqrt{C_{0} |\tilde{C_{00}}|}}{2} \, t_{i} - |t_{0}| \big) = e^{C_{0}N}  (1 + \frac{C_{0}} {2}) -  1  $. 
Evaluating the value of $ \epsilon_{1} $ for 60 e-folding number one can see that 
\be 
\epsilon_{1} = \frac{C_{0}}{2 (e^{C_{0}N} (1 + \frac{C_{0}}{2}) - 1)}  \ll 1. 
\ee

The time instance for which the inflation ends corresponds to the condition $\epsilon_{1} = 1 $ \cite{OO_2021} and references therein. This gives $ \sinh^{2}\big(\frac{\sqrt{C_{0} |\tilde{C_{00}}|}}{2} \, t_{f} - |t_{0}| \big) = \dfrac{C_{0} }{2} \ll 1 $ where 
\be
t_{f} = \frac{2 \big(|t_{0}| + \ln(\sqrt{1 + \frac{C_{0}}{2}} - \sqrt{\frac{C_{0}}{2}}) \big)}{\sqrt{C_{0} |\tilde{C_{00}}|}}
\ee
is the time of end of inflation and it is obvious that $ t_{f} < t_{cr} $. It is worth briefly noting that in eq. (64) there is a $ \pm $ sign under the logarithm. However to satisfy the requirement $  t_{i} <  t_{f} $ the minus sign must be chosen. Then this requirement reduces to $ C_{0}N > 0 $ and is satisfied automatically for $ C_{0} > 0 $. 

Consider now the Hubble radius $ R_{H} =  \frac{1}{\dot{a}} $. Because $ t_{f} $ is exactly the time moment for which the Hubble radius starts to increase let us look at the behaviour of function $ R_{H}  $ for the time domain $ t < t_{cr} $. Solving $ \dot{R_{H}} = 0 $ we obtain the minimum just at $ t =  t_{f} $. Such a fact may indicate that the model under consideration describes the desired behaviour of inflationary evolution and can be considered as a completely possible inflationary model \cite{OO_2021} and references therein.  

Now one need to consider the remaining slow-roll indices $ \epsilon_{3} $ and $\epsilon_{4} $. Let us express the index $ \epsilon_{3} = \frac{\dot{F_{R}}}{2 H F_{R}} $  through the variable $ z = \frac{-C_{0} (R - 3 \tilde{C_{00}})}{3 \tilde{C_{00}} (C_{0} + 4)} = \tanh^{2}\big(\frac{\sqrt{C_{0} |\tilde{C_{00}}|}}{2} \, t - |t_{0}| \big) < 1 $ and the function 
\be
F(R) = C \,\, {_{2}F_{1}} (\alpha; \beta; -\frac{1}{2}; z) + \tilde{C} \,\, z^{\frac{3}{2}} \,\, {_{2}F_{1}} (\alpha + \frac{3}{2}; \beta + \frac{3}{2}; \frac{5}{2}; z)  
\ee
where $ \alpha = - k_{\pm} $, $ \beta = - k_{\mp} $, see eq. (19). Then 
\be
\epsilon_{3} = - \frac{C_{0}}{2 \cosh^{2}\big(\frac{\sqrt{C_{0} |\tilde{C_{00}}|}}{2} \, t - |t_{0}| \big) } \frac{F_{zz}}{F_{z}}
\ee
and it is necessary to show that can be $ \epsilon_{3} \ll 1 $. First we put $ \tilde{C} = 0 $ and show that function $ \frac{F_{zz}}{F_{z}} $ is monotonic. So from $  \frac{d}{dz} \frac{F_{zz}}{F_{z}} = 0 $, $ F_{z} \neq 0 $, follows that $ F = a e^{bz} + c $ and this is not the hypergeometric  function. Then one need to understand whether $ \frac{F_{zz}}{F_{z}} $ decreases or increases. Because $ z \in [0, 1) $ we find the value of this function at $ z = 0 $ and $ z \longrightarrow 1$. At point $ z=0 $ the hypergeometric function is represented by series: $ {_{2}F_{1}} (\alpha; \beta; \gamma; z)  = \sum\limits_{n=0}^\infty \frac{\Gamma(\alpha+n)\Gamma(\beta+n)\Gamma(\gamma)}{\Gamma(\alpha)\Gamma(\beta)\Gamma(\gamma+n) }  \frac{z^n}{n!}   $ ($ \Gamma(*)$ is Gamma function) and we obtain 
\be
\frac{F_{zz}}{F_{z}} \Big|_{z=0} = 2(\alpha + 1)(\beta + 1) = \frac{2}{C_{0}}. 
\ee
In order to find the value of $ \frac{F_{zz}}{F_{z}} $ at $ z \longrightarrow 1$ we write the hypergeometric function in the form: $ {_{2}F_{1}} (\alpha; \beta; - \frac{1}{2}; z)  =  \tilde{\Gamma}  {_{2}F_{1}} (\alpha; \beta; \alpha + \beta + \frac{3}{2}; 1 - z) + \tilde{\tilde{\Gamma}} (1-z)^{- \alpha - \beta - \frac{1}{2}}  {_{2}F_{1}} (- \alpha - \frac{1}{2}; - \beta - \frac{1}{2}; - \alpha -  \beta + \frac{1}{2}; 1 - z)  $, $ \tilde{\Gamma} $ and $ \tilde{\tilde{\Gamma}}  $ are constants expressed through Gamma functions of $ \alpha $ and $ \beta $ \cite{NU}. Representing both hypergeometric functions in the form of series and writing the exponent of $ z $ as $1 + \frac{1}{C_{0}} $ we find that the second term is equal to zero. Then
\be
\frac{F_{zz}}{F_{z}} \Big|_{z \longrightarrow 1} = - \frac{(\alpha + 1)(\beta + 1)}{\alpha + \beta + \frac{5}{2}} = \frac{1}{1 - C_{0}} 
\ee
Thus one can see that the function $ \frac{F_{zz}}{F_{z}} $ monotonically decreases. For the time $ t = t_{1} $ we obtain
$ z_{i} = 1 -  e^{ -(C_{0}N + \ln(1 + \frac{C_{0} }{2}) ) } \sim 1 -  e^{ -C_{0}N} (1 - \frac{C_{0}}{2}) $, $ \cosh^{2}\big(\frac{\sqrt{C_{0} |\tilde{C_{00}}|}}{2} \, t_{i} - |t_{0}| \big) = e^{C_{0}N}  (1 + \frac{C_{0}} {2}) $ and the requirement $ \epsilon_{3}(z_{i}) \ll  1 $ can be satisfied by choosing a sufficiently small $ C_{0} $. So using a direct calculation one can obtain  that for $ C_{0} = 0.001 $ the index $ \epsilon_{3} \cong - 0.00796 $. When one consider the complete solution (68) it is impossible to establish the behaviour of the function $ \frac{F_{zz}}{F_{z}} $ so simply because at $ z\longrightarrow 0  $ $ F_{z} = 0 $ and $ F_{zz} \nrightarrow $; it can be shown that $ \frac{F_{zz}}{F_{z}} \longrightarrow \pm \infty$ for $ \tilde{C} \lessgtr 0 $. Considering that at $z \longrightarrow 1 $ the function $ \frac{F_{zz}}{F_{z}} $ has a finite positive value one can conclude that for $ \tilde{C} > 0 $ an extremum exists. However we will not analyse this further since the possibility of doing $ \epsilon_{3} \ll 1 $ has already been shown.

Similar to the previous one expressing the slow-roll index $\epsilon_{4} = \dfrac{\ddot{F_{R}}}{H \dot{F_{R}}} $ through the variable $ z $ we obtain
\be
\epsilon_{4} = - \frac{C_{0}}{2 \sinh^{2}\big(\frac{\sqrt{C_{0} |\tilde{C_{00}}|}}{2} \, t - |t_{0}| \big) } + \frac{C_{0}}{2} - \frac{C_{0}}{\cosh^{2}\big(\frac{\sqrt{C_{0} |\tilde{C_{00}}|}}{2} \, t - |t_{0}| \big) } \frac{F_{zzz}}{F_{zz}}.
\ee
The first two terms are much less than one and we study the third term as for $ \epsilon_{3} $. Putting $ \tilde{C} = 0 $ we obtain that $ \frac{F_{zzz}}{F_{zz}} $ is monotonic and 
\ba 
\frac{F_{zzz}}{F_{zz}} \Big|_{z=0} & = & \frac{2}{3}(\alpha + 2)(\beta + 2) = 1 \nonumber \\
\frac{F_{zzz}}{F_{zz}} \Big|_{z \longrightarrow 1} & = & - \frac{(\alpha + 2)(\beta + 2)}{\alpha + \beta + \frac{7}{2}} =  \frac{3 C_{0}}{2(1 - 2 C_{0})}, 
\ea
whence it follows that $ \epsilon_{4} $ can also be made much less than one. Same as $  \epsilon_{3} $ using a direct calculation we obtain that for $ C_{0} = 0.001 $ the index $ \epsilon_{4} \cong - 0.00754 $. 

To estimate to a few decimal places the values of observational inflation indices namely the spectral index of primordial scalar perturbations, the tensor-to-scalar ratio and the spectral index of tensor perturbations \cite{NOO}, \cite{HN}, \cite{K}, \cite{OO_2021}, expressed in terms of the slow-roll indices we cannot give a satisfactory analytical expression. By direct calculation for $ C_{0} = 0.001 $ we find:
\ba
n_{s} & = & 1 - \frac{4 \epsilon_{1} - 2 \epsilon_{3} + 2 \epsilon_{4}}{1 - \epsilon_{1 }} \cong 0.966826, \,\,\, 
r = \frac{48 {\epsilon_{3}}^2}{( 1 + \epsilon_{3})^2} \cong 0.0030911 \nonumber \\
n_{T} & = & -2 ( \epsilon_{1} + \epsilon_{3})  \cong  -0.000112105.
\ea
From this one can see that the values of these quantities are of the same order and are very close to the values obtained  in the work \cite{OO_2021}. This may indicate that the inflation phenomenology of the model under consideration in the domain of its definition is close to the $ R^2 $ model and the deformed model \cite{OO_2021}.

\section{Conclusion}

In this paper we obtain a number of exact solutions of the Friedmann cosmological equation for metric $ F(R) $ gravity. We focus on the possibility of obtaining exact dependences on cosmological time simultaneously for the most interesting physical quantities, such as the scale factor, scalar curvature, Hubble rate and function $ F(R) $ itself. It was not possible to find $ F(R) $ exactly everywhere, but only in the form of a series, but the priority was the ability to express $ a(t) $ through $ t $. Some of the presented solutions were obtained earlier in the cited works. In the same works other solutions were obtained for which we can represent function $ \Phi(H^{2}) $, but at the same time it is not possible to express the scale factor, the e-folding number or (and) $ H^{2} $ through the time coordinate; so we did not consider them. 

We also carried out a physical analysis of one of the solutions obtained and found that this model has the adequate cosmological interpretation. So this solution describes a model with an acceptable inflation phenomenology.

\end{document}